\documentclass[preprint]{acmart}

\usepackage{color}
\usepackage{graphicx}
\usepackage{subcaption}
\usepackage{caption}
\usepackage{tcolorbox}
\usepackage{hyperref}
\usepackage{listings}
\usepackage[LGR,T1]{fontenc}
\usepackage[utf8]{inputenc}
\usepackage{textalpha}
\newcommand\Que[1]{%
   \leavevmode\par
   \stepcounter{question}
   \noindent
   \thequestion. Q --- #1\par}
\renewcommand\footnotetextcopyrightpermission[1]{} 
\begin{document}

\title{Making Sense of Data in the Wild: Data Analysis Automation at Scale} 


\author{Mara Graziani}
\affiliation{
  \institution{IBM Research Europe}
  \streetaddress{4 Saümerstrasse, Rüschlikon}
  \country{Switzerland}
}
\affiliation{
  \institution{National Center for Competence in Research - Catalysis}
  \country{Switzerland}
}
\email{mara.graziani@ibm.com}

\author{Malina Molnar}
\affiliation{
  \institution{IBM Research Europe}
  \streetaddress{4 Saümerstrasse, Rüschlikon}
  \country{Switzerland}
}
\affiliation{
  \institution{National Center for Competence in Research - Catalysis}
  \country{Switzerland}
}

\author{Irina Espejo Morales}
\affiliation{%
  \institution{IBM Research Europe}
  \country{Switzerland}
}
\author{Joris Cadow-Gossweiler}
\affiliation{%
  \institution{IBM Research Europe}
  \country{Switzerland}
}

\author{Teodoro Laino}
\affiliation{%
  \institution{IBM Research Europe}
  \country{Switzerland}
}
\affiliation{
  \institution{National Center for Competence in Research - Catalysis}
  \country{Switzerland}
}

\renewcommand{\shortauthors}{Graziani et al.}

\begin{abstract}
As the volume of publicly available data continues to grow, researchers face the challenge of limited diversity in benchmarking  machine learning tasks. 
Although thousands of datasets are available in public repositories, the sheer abundance often complicates the search for suitable data, leaving many valuable datasets underexplored. 
This situation is further amplified by the fact that, despite longstanding advocacy for improving data curation quality, current solutions remain prohibitively time-consuming and resource-intensive. 
In this paper, we propose a novel approach that combines intelligent agents with retrieval augmented generation to automate data analysis, dataset curation and indexing at scale. 
Our system leverages multiple agents to analyze raw, unstructured data across public repositories, generating dataset reports and interactive visual indexes that can be easily explored. 
We demonstrate that our approach results in more detailed dataset descriptions, higher hit rates and greater diversity in dataset retrieval tasks. 
Additionally, we show that the dataset reports generated by our method can be leveraged by other machine learning models to improve the performance on specific tasks, such as improving the accuracy and realism of synthetic data generation. 
By streamlining the process of transforming raw data into machine-learning-ready datasets, our approach enables researchers to better utilize existing data resources.
\end{abstract}

\begin{teaserfigure}
  \includegraphics[width=\textwidth]{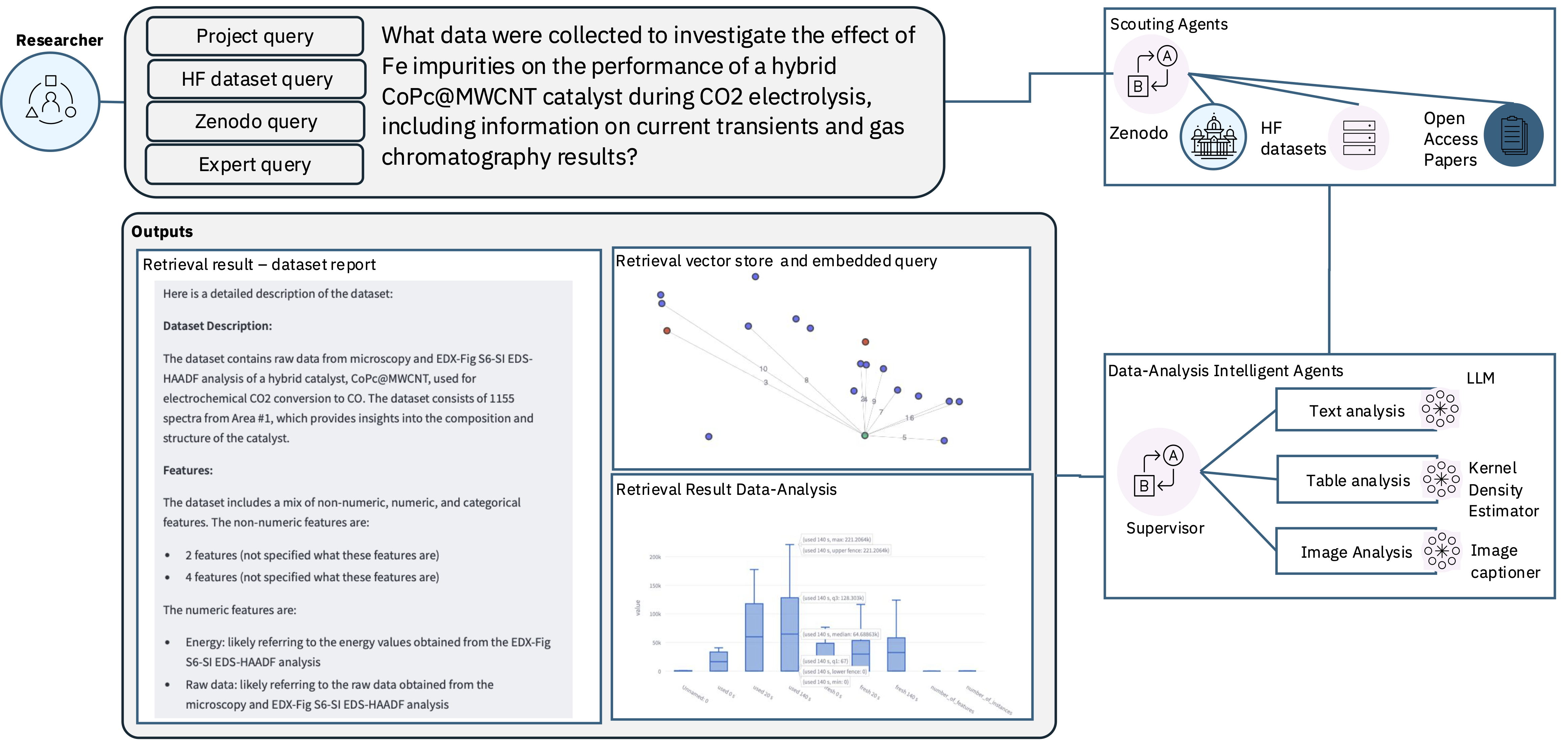}
  \caption{Overview of the proposed system to retrieve relevant datasets based on user queries.}
  \label{fig:teaser}
\end{teaserfigure}

\received{ XXX}
\received[revised]{YYY}
\received[accepted]{ZZZ}

\maketitle

\pagestyle{plain} 
\section{Introduction}
In a world where digitization is generating head-spinning volumes of data, one may wonder why large part of machine learning research is confined to a handful of benchmarks~\cite{koch2021reduced}.
As of June 2024, Zenodo's repository exceeds one petabyte of data, hosting over four million records that encompass all sorts of research outputs~\cite{zenodo_stats}. 
In a similar fashion, Hugging Face (HF) datasets offers more than 250 thousand datasets that are ready-to-use for machine learning purposes~\cite{lhoest-etal-2021-datasets}.
Governmental funding agencies increasingly mandate the open publication of all scientific results to promote transparency and reuse\cite{wilkinson2016fair,hodson2018turning}. 
However, scientists often lack the domain knowledge or resources to curate and document their data in a way that ensures reusability across disciplines. 
This gap limits the effective utilization of newly available datasets, contributing to the underuse of the vast volume of created knowledge, making it harder to identify and select relevant data~\cite{schwartz2015paradox}. 
As a result, researchers gravitate toward easy-choice datasets, such as those with well-documented and extensively explored content~\cite{hemphill2022properties,sheridan2021datadiscovery}, even for tasks beyond their original intended purpose~\cite{koch2021reduced}. 
This tendency reinforces the underutilization of less curated, yet potentially valuable, datasets and ultimately limits scientific discovery.

It is well established that high-quality data curation enhances the findability and reusability of datasets~\cite{hemphill2022properties,sheridan2021datadiscovery}. 
However, curation tasks such as creating data catalogs~\cite{sheridan2021datadiscovery}, maintaining structured databases~\cite{mysql} and preparing data sheets~\cite{gebru2021datasheets} are time-consuming, expensive and rarely performed by researchers themselves.
In contrast, sharing raw and unstructured data is a more practical approach, enabling researchers to focus on their core work.
To combine the benefits of data curation with the practicality of raw data sharing, one needs a system that can automatically handle curation and indexing. 

We hypothesize that datasets can be organized at scale by combining the capabilities of intelligent agents with Retrieval-Augmented Generation (RAG)~\cite{lewis2020retrieval}. 
Unlike existing methods, which rely on manual or semi-automated processes to analyze single files~\cite{jaimovitch2023can,ansari2024agent,cheng2023gpt}, our method employs multiple agents to automate the data analysis of the raw, unstructured data across entire portions of online public repositories. 
This enables the generation of an index that can be visually explored and interacted with through augmented generation. 
While our goal is to provide a proof-of-concept, our approach is already highly generalizable to files uploaded without structured curation, ranging from public repositories like Zenodo~\cite{zenodo_communities_page} to the extensive collection of machine-learning-ready datasets on HF datasets~\cite{lhoest-etal-2021-datasets}. 
By interactive visualization of the data analysis results and of data examples, the task of the researchers is simplified to the exploration of an interactive web application, facilitating a seamless transition from experimental data to machine learning-ready data, and enhancing the overall efficiency of scientific research. 
Moreover, as synthetic data generation becomes a reality~\cite{villalobos2024position,shumailov2024ai}, we demonstrate that the analysis generated by our approach can improve the quality and reliability of data synthetization results, enabling researchers to fully exploit the content of existing data repositories in the wild. 
\section{Methods}
\label{sec:methods}
As of now, the process of finding datasets evolves around the following steps: (i) search for publications related to the topic of interest; (ii) identify the related open-access datasets; (iii) download one or more datasets; (iv) perform data wrangling and analysis; and (v) re-iterate on the results to find additional similar or related datasets. Each of these steps can be daunting and tedious, requiring a combination of serendipity and domain-specific expertise.

A system that automates data curation should be able to perform these steps, and fulfill the following requirements in terms of functionalities:
\begin{enumerate}
\item find datasets matching queries in natural language;
\item automate data analysis; 
\item generate visualizations that facilitate further inspections;  
\end{enumerate}
To this end, we propose a data-scouter system that integrates a variety of advanced machine learning models, including intelligent assistants implementing the ReACT (Reasoning and Acting) framework provided by Langchain\cite{langchain} and automated execution pipelines. 

In the following sections, we describe in detail the different modules. For instance, Section \ref{sec:scraping-zenodo} introduces the agents in charge of downloading the data and finding relevant publications, hence solving tasks (i) to (iii). Sections~\ref{sec:file-analysis} explain the methods implemented for automating the analysis of the data and addressing requisite (2) by performing step (iv).
Focusing on requisites (1) and (3), Sections~\ref{sec:RAG-indexing} and~\ref{sec:interactive-vis} provide the details of the implementation of the RAG indexing and the interactive web-based visualization. 
Finally, Section~\ref{sec:evaluation} gives an overview of the criteria and measures used to evaluate the performance of our approach as opposed to the existing solutions. 

\subsection{Download of Zenodo Community Data}
\label{sec:scraping-zenodo}
A data \textit{community} in Zenodo is a curated collection of publications, datasets, software and other research outputs that is identifiable through a tag. 
Each community is focused on a specific research area, allowing contributors to group related works and funders to track tangible results of the funding. 
As of November 2024, Zenodo lists $6176$ communities, of which $3025$ correspond to projects, $1945$ to  organizations, $749$ to topics, and $448$ to events~\cite{zenodo_communities_page}. 

In the proposed system, two data downloading agents work together to systematically collect raw data from Zenodo. 
Note, every single downloaded record underwent a rigorous governance check to ensure compliance with licensing requirements. 
This process, performed offline, verifies that the license under which the data is released permits its use in alignment with ethical and legal standards. 
This critical validation process is essential for advancing toward a world in which AI development adheres to principles of responsibility, transparency, and legality, ensuring that no datasets are used in violation of licensing agreements or broader legal frameworks.

The first agent downloads file records via API integration and HTTP requests. 
By leveraging the Zenodo REST APIs for accessing community endpoints, the agent gains direct access to all records associated with a specific community. 
These API requests require an access token, which is included in the request header.
The second agent is tasked with retrieving open-access publications linked to the records. 
This is achieved by simulating user interactions with the browser through the Playwright automation toolkit~\cite{playwrightbrowsertoolkit}. 
Developed by Microsoft and integrated into Langchain, Playwright allows us to emulate the navigation to a publication's Discriminative Object Identifier (DOI) and the consequent download of the article if it is open-access. 
The conceptual workflow follows three main steps. 
First, the agent identifies the download URL from the record webpage. By querying via REST API the Unpaywall~\cite{unpaywall} service, the agent verifies if the publication is open-access, and if so it downloads it. 
If by any reason the service provides no usable response, the agent follows the DOI landing page and searches for any HTML meta tags linking to a publication. 

In addition to the dataset files and the publication, this module stores all the metadata available on the original record page, including the user-generated description of the datasets. This field will then be used to compare our generated results to the original descriptions. 

\subsection{Multi-modal file analysis}
\label{sec:file-analysis}
Our system utilizes multiple AI models to enhance traditional data analysis techniques and extend their applicability to different modalities. 
The integration with AI models is designed to be flexible and adaptable to the rapidly evolving field, supporting a wide range of large language models (LLMs). Specifically, we have integrated proprietary models~\cite{mishra2024granite}, open-source models such as Llama 8B and 70B~\cite{touvron2023llama}, and CPU-based small parameter models such as SmolLM-135M~\cite{smollm}. 
A step-by-step description of the process is articulated in the following. 

\subsubsection{Data Loading} 
\label{sec:data-loading}
Multiple agents run in parallel to analyze the files based on their extensions. Dedicated dataloaders are used to load the data in common formats for documents (i.e. PDF), tabular data (i.e. CSV and XML), images (i.e., JPG, PNG, and TIFF) and raw text (i.e. TXT). For unsupported formats, users can implement custom loaders using Python code. 
Once the files are loaded, the data are transformed into the HuggingFace dataset format for analysis. 
At this stage, the first analysis performed involves the detection of the modality and type of each feature in the data. 

\subsubsection{Modality-specific file analysis} 
The data analysis of each file is conducted by modality-specific agents that merge traditional data analysis techniques with AI models. 
For example, the table analyzer utilizes a uni-modal Kernel Density Estimation (KDE) for each feature and calculates feature correlations and feature predictability.
The image analyzer applies image captioning models to convert image content to textual descriptions.
Lastly, the text analyzer summarizes long textual data, such as the papers, in bullet points notes. 
This analyzer also implements word counts analyses to detect topics and word distributions over the vocabulary. 
While already covering multiple modalities and formats, our approach is modular and can thus be easily extended to additional modalities by integrating new agents. 

\subsubsection{Centralized file-level aggregation}
The results from the modality-specific analyzers are sent to a central variable aggregation system that converts each result into a string with hash indexing. 
This method facilitates efficient storage and retrieval of the results from each agent, ensuring all relevant information is readily accessible for further processing. 
An LLM-based supervisor agent then analyzes the results and through a map-reduce prompting strategy obtains a data content summary covering information on the feature names and descriptions, KDE estimations, publication key points and textual word distributions over the vocabulary. 
A second prompt is then used to ask the supervisor to generate an overarching description of the file, based on the data content summary. 
The overarching description prompt is the following:
\begin{tcolorbox}
"\texttt{You are a helpful data analyst. You have been given metadata about a dataset. Write a short description for the dataset. Consider the following additional information [SUPERVISOR-OUTPUT]. If possible from the labels, say what the data content is.}"
\end{tcolorbox}
In the prompt, the \texttt{[SUPERVISOR-OUTPUT]} placeholder is dynamically replaced with the data content summary generated by the supervisor. 

\subsubsection{Extension to HF datasets}
Since the data loading step in Section~\ref{sec:data-loading} involves the conversion of the data to the HF datasets format, our data analysis system naturally extends to the analysis of HF datasets. 
An additional tool enables the download of HF datasets from the name identifier and specified parameters. 
The interfacing is crucial for making sense of data in the wild, as it allows for a direct comparison of data repositories in Zenodo to large-scale, curated and ML-ready datasets. 
By doing so, we can identify similarities and alternative datasets to an individual Zenodo record, ultimately enhancing the utility and applicability of the retrieval. 

\subsubsection{Hierarchical record-level report}
\label{sec:hierarchical-reports}
Each dataset record in a Zenodo upload comprises an arbitrary number of files. 
The outputs of the multimodal file analysis step include as many file-level reports as the supported files in the record. 
Each file-level report offers a comprehensive description of the content within each file. 
To create a hierarchical report that describes the dataset in its entirety, another map-reduce agent is utilized to aggregate information at the dataset level. 
The file-level descriptions are mapped to individual summaries, which are then consolidated into a single, unified summary.

\subsection{RAG indexing}
\label{sec:RAG-indexing}
The generated file-level and record-level reports are used in conjunction with the original user descriptions to create a comprehensive and easily searchable representation of all the analyzed records. The RAG indexing process involves concatenating the LLM-description and the user-descriptions, dividing the text into chunks, and creating a vector embedding for each chunk using a transformer-based model. A global vector representation is then obtained through averaging the individual vectors. This vector representation captures the semantic meaning and relationships between the different components of the description, enabling efficient and accurate search and retrieval. The vector encoding generated by the transformer-based model is then used for the indexing and the augmented retrieval generation and to answer user queries. 

\subsection{Interactive visualization}
\label{sec:interactive-vis}
The interactive web-based visualization application was designed to accommodate a broad spectrum of users, regardless of their technical expertise. 
Built on Streamlit~\cite{streamlit}, it enables users to retrieve datasets and gain meaningful insights about the data without needing to write any code. 
The user is requested to enter a natural language query and the app instantly generates a graph-based representation of the available records within the community. 
In the graph, each record is depicted as a node, with edges connecting the nodes to indicate the relevance of each record to the query.
The relative positioning of the nodes is determined by the Fruchterman-Reingold force-directed algorithm~\cite{fruchterman1991graph}, for which an implementation is available in NetworkX~\cite{networkx}. 
In the algorithm, nodes are attracted to each other based on the vector similarity of their descriptions. 
Particularly, nodes with high similarity to others are positioned so to form an agglomeration cluster, while dissimilar nodes appear isolated. 
This visual representation highlights the relationships and differences between records, enabling users to quickly chain from one record to the next relevant one.

\subsection{Evaluation}
\label{sec:evaluation}
In evaluating our system, we simulate the retrieval of relevant datasets by a researcher operating for a specific use case. 
Starting from open-access contributions with Zenodo traceability, we base our evaluation on several components, which are outlined in the following subsections. 

\subsubsection{Qualitative assessment} 
We start by comparing generated and pre-existing descriptions from a qualitative standpoint.
To evaluate the impact of diverse metadata fields, we compare the results obtained by using only the original description as metadata, and by then augmenting this information with the paper content, feature descriptions and a few dataset examples. 
Remaining on the line of qualitative assessment, we extended the analysis to all the files in the community by evaluating the length of the generated descriptions and their coherence with the pre-existing ones. 
While longer descriptions may not necessarily mean a better result, we see from several examples (see Appendix~\ref{appendix:qualitative-assessment}) that, in these cases, the additional words are used to provide a comprehensive and detailed overview of the dataset. 
\subsubsection{Redundancy of pre-existing descriptions}
We evaluate the correlation between the pre-existing descriptions and the corresponding papers to assess the relevance and usefulness of the descriptions.
This assessment is performed at the representation level in the vector store. 
For each dataset entry in the community, we compute the cosine similarity between the vector representing the original description and that of the related paper.
We also compute the cosine similarities between the generated descriptions and the same paper. 
If the information in the original descriptions is strongly redundant with the papers, we would then expect the distribution of the absolute values of the cosine similarities to be shifted towards one. 
\subsubsection{Question retrieval accuracy}
To measure the effectiveness of our question-answering capabilities, we perform a retrieval experiment. 
A LLM is used to generate questions about the dataset files starting from the paper information. 
The prompt to generate the questions is explicitly formatted so that the questions that are generated can only be answered by having access to the data and knowing the content of the files.
Since each of these questions is associated to the starting record, we then evaluate the performance of the RAG agent in re-identifying the starting papers. 
For transparency, we report the prompt in the following: 

\begin{tcolorbox}
"\texttt{Here you have the summary of a paper about a dataset: \{text\}. Based on the summary of the paper, create a list of 15 questions that include enough information to understand what is the paper, but that would require access to the data used for the experiments in the paper to be answered. The goal of the question is to identify a dataset that could bring to the discoveries made in the paper.}"
\end{tcolorbox}

To evaluate this experiment, we start by measuring the rate of exact matches, which corresponds to the top-1 accuracy of identifying the starting record id from each question. 
We then relax this evaluation to top-5 and top-10 accuracy by including the first 5 and 10 documents retrieved, respectively. 
Since more than one dataset may contribute to answering a specific question, we also evaluate the \textit{diversity} of the retrieval results., which is defined as the normalized entropy of the top 10 retrieved results. 
A low entropy for a successfully retrieved document (top-1 hit) means that the question was relatively easy to solve. 
However, this also shows that the results of the retrieval fails to give more insights of what other datasets may be related to each other. Conversely, high entropy means that there is high uncertainty around the retrieval results and that the question is probably unclear or that the retrieval results are more likely to be polluted by unrelated datasets. 
Therefore, ideal normalized entropy scores should be distributed far away from these two extremes. 

\subsubsection{HuggingFace integration} 
\label{evaluation-hf}
To test the integration with HF, we compare the Zenodo datasets with a set of known, similar datasets from the HF datasets repository. 
We compute the vector cosine similarity between a Zenodo record generated description and the generated description for a related HF dataset, assessing that the similarity is higher than that between to unrelated datasets. 

\subsubsection{Re-usability of data analysis results}
We demonstrate the value of the insights in the file-level reports by assessing their potential to enhance machine learning model performance on a particularly challenging task: generating realist synthetic data samples. 
Specifically, we use the file-level reports to inform a synthetic data generation pipeline, similar to the approach outlined in~\cite{xu2024llms}, as opposed to relying solely on dataset examples~\cite{borisovlanguage}. 
While our goal is not to compete with state-of-the-art data synthesis techniques, this evaluation provides an opportunity to measure the downstream impact of the file-level reports generated by the agents. 
We explore how the information gained by the data analysis effects the quality of synthetic data, focusing on how closely the generated data aligns with the distribution of the original dataset. 

To generate the synthetic samples, we query an agent with access to a Python interpreter tool (PythonREPL)~\cite{langchain}. 
The prompt incorporates the information in the file-level report and instructs the agent to generate Python code to produce synthetic data. 
The agent then evaluates and iteratively refines the script. 
This approach proved to be the most flexible for our evaluation, as it is both general and adaptable. 
For instance, it allowed us to compare the results generated with and without access to the file-level reports with minimal changes between the two conditions. 
For more details, Appendix~\ref{appendix:prompt_data_synth} reports the prompts used for the experiments and Appendix~\ref{appendix:scripts_data_synth} presents the code scripts generated by the agent. 
Both experiments have access to 5 original examples. The base chat LLM used is~\textsc{meta-llama/llama-3-1-70b-instruct} ~\cite{llama31}, for more details about generation parameters see Appendix~\ref{appendix:params_data_synth}.

\section{Results}
\begin{figure}[t]
    \centering
    \includegraphics[width=\linewidth]{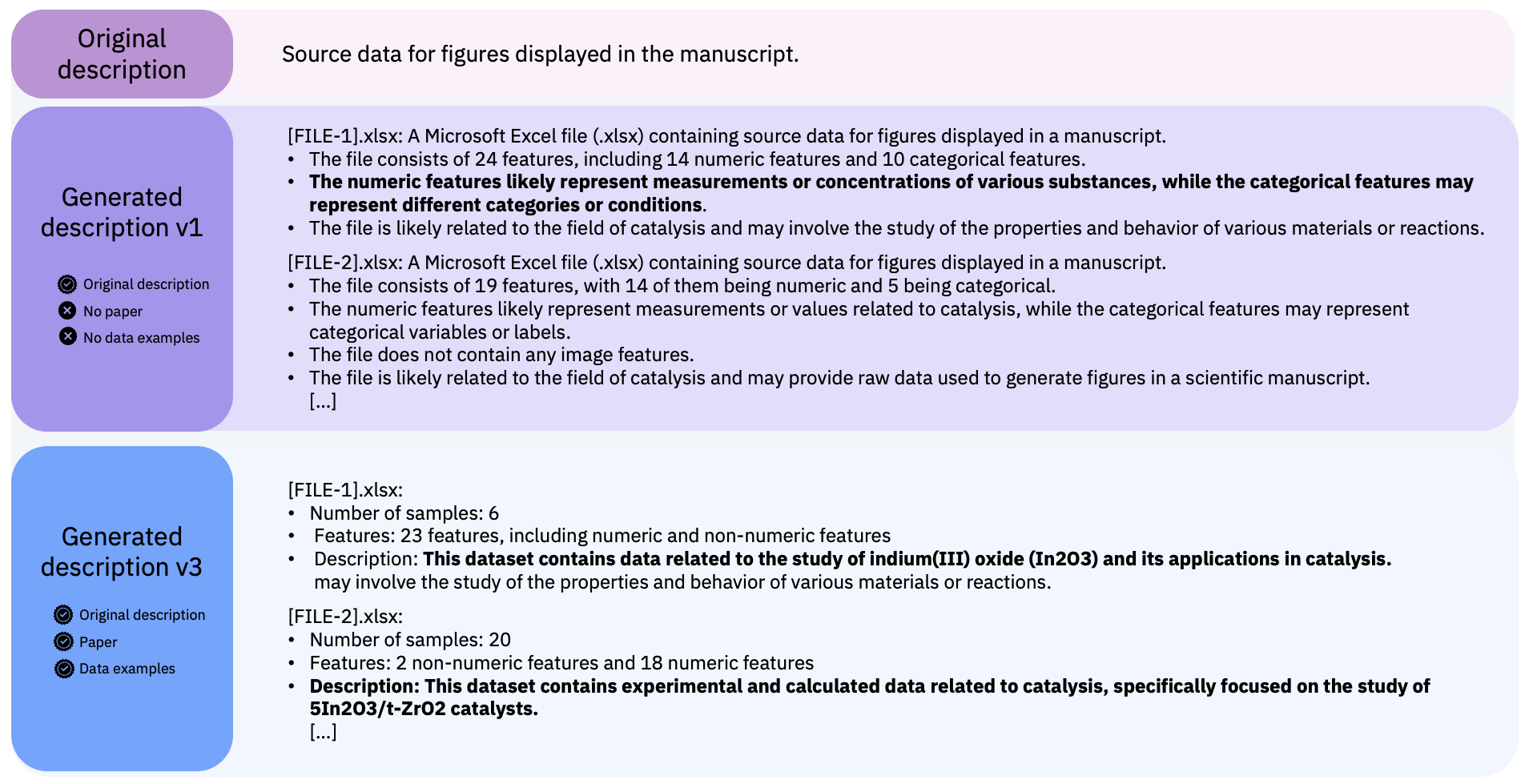}
    \caption{A comparison of a user-provided description that lacks informativeness for data reuse with our generated descriptions: V1, created solely from data analysis results, and V3, which incorporates data analysis results, the paper, and data examples.}
    \label{fig:examples}
\end{figure}

\begin{figure*}[!ht]
\centering
\subfloat[]
{\includegraphics[width=0.3\linewidth]{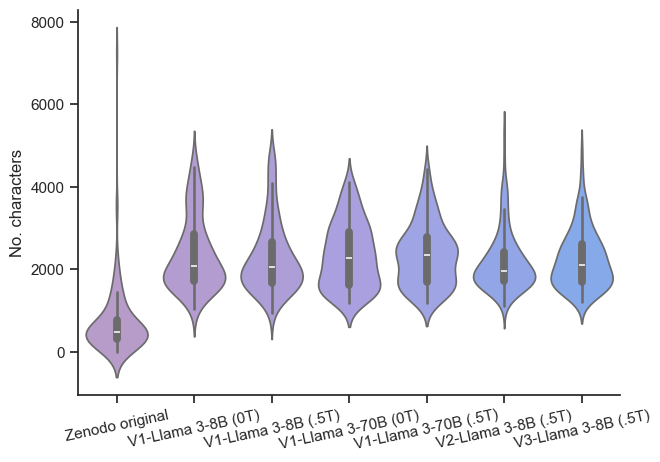}}
\subfloat[]
{\includegraphics[width=0.3\linewidth]{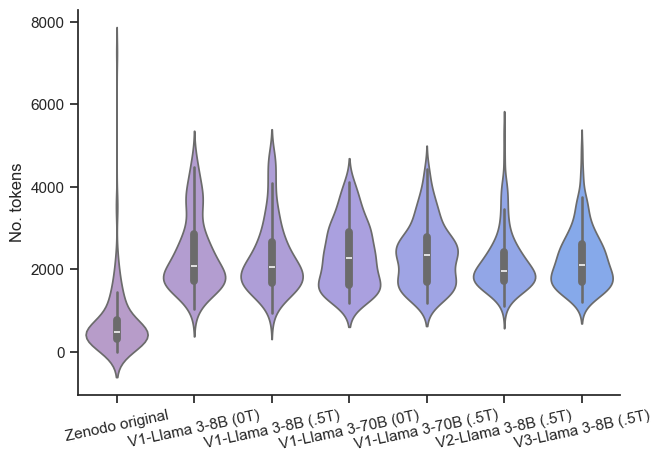}}
\subfloat[]
{\includegraphics[width=0.3\linewidth]{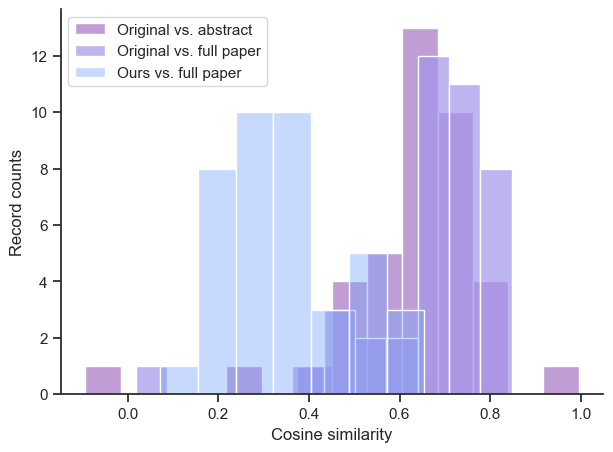}}
\caption{(a, b) Comparison of user-defined (original) and generated description lengths, measured in (a) number of characters and (b) number of tokens. (c) Distribution of pairwise similarities: original descriptions compared to their corresponding papers versus generated descriptions compared to the same papers.}
\label{fig:length}
\end{figure*}
The Zenodo community used in the experiments is named \textit{nccr-catalysis} and focuses on the problem of accelerating chemical reactions via catalysts, substances that lower the activation energy without being consumed in the process~\cite{6641837dataset,6381537dataset,12914140dataset,6401829dataset,12683880dataset,6319808dataset,8082748dataset,8239352dataset,8075771dataset,8246660dataset,6325256dataset,7199360dataset,6631389dataset,8116794dataset,8178921dataset,8068699dataset,8134490dataset,4652040dataset,5665271dataset,10410523dataset,10033139dataset,8074143dataset,6381625dataset,8099440dataset,8273259dataset,8082205dataset,4043189dataset,8229210dataset,8250270dataset,8229079dataset,10048201dataset,8073393dataset,8239023dataset,6865680dataset,4046256dataset,5931544dataset,8246384dataset,8019944dataset,6634788dataset,8271972dataset,6451293dataset,13134128dataset,8085800dataset,6547649dataset,6193067dataset,6659731dataset,7656889dataset,10351719dataset,6511235dataset,6421142dataset,6359849dataset,12165448dataset,8315184dataset,8183015dataset,7274529dataset,6325397dataset,8134226dataset,10022311dataset,7107186dataset,8027571dataset,6350245dataset,8248120dataset,6840454dataset,7380563dataset,8116584dataset,10050308dataset,6908494dataset,7675514dataset,6135989dataset,6786359dataset,8095929dataset,8029087dataset,8119767dataset,6037503dataset,8147183dataset,8246816dataset,8271809dataset,6720368dataset,6340467dataset,7380519dataset,10184425dataset,6482769dataset,8228815dataset,7359881dataset,8085998dataset,7991566dataset,5723297dataset,6752507dataset,8152643dataset,8271889dataset,5848181dataset,8272032dataset,8271015dataset,10161171dataset,6541445dataset,8037077dataset,8099016dataset,8402733dataset,8036937dataset,8374365dataset,8399256dataset,8134171dataset,7707974dataset,10620282dataset,7849574dataset,8074911dataset,10524037dataset,7117080dataset,8276926dataset,8055442dataset,8028644dataset,12515595dataset,13771489dataset,8051277dataset,12578395dataset,6259193dataset,8271785dataset,5767223dataset,8086271dataset,11186147dataset,6380116dataset,5205584dataset,6511898dataset,8366598dataset,8246442dataset,7446294dataset,8167572dataset,8420551dataset,8390247dataset,8305660dataset,8074042dataset,8246472dataset,6325285dataset,8119797dataset,7002917dataset,12104162dataset,vavra2022cu}. This community was created following a mandate by the Swiss National Science Foundation to make all data funded through the NCCR program publicly available. It serves as a clear example of a dataset community containing state-of-the-art catalysis research data and led by researchers who are experts in their scientific domains but often lack expertise in AI or data curation. As a result, many datasets in this community, and similar ones on Zenodo, frequently lack sufficient metadata or documentation, limiting their reusability in contexts beyond their original purpose.
The data related to catalysis is diverse, multidisciplinary and multimodal, spanning multiple domains such as chemistry, physics, and computational sciences.
Catalytic systems identification often involves extensive experimentation efforts that may not be consistently digitized, leading to a lack of large-scale, structured datasets. In this context, improving data quality is crucial for enhancing the reusability and retrievability of the datasets.

Where not specified otherwise, the models used for the experiments are Llama-8B~\cite{touvron2023llama}, and sentence transformers (all-mpnet V2)~\cite{reimers-2019-sentence-bert} to obtain the vector encoding. 

\subsection{Description quality on ablations}

Figure~\ref{fig:examples} compares the quality of the descriptions generated by passing different information to the language model. 
The example is representative of a case where Zenodo users provided little to no information about the data. 
Despite this, our method is able to identify relevant features and generate meaningful descriptions. 
By comparing the description generated in version 1 (where no paper information nor data examples were provided to the model) with those generated in version 3 (where both paper and data examples were used), we observe that incorporating these additional resources improves the description quality. 
Specifically, using the paper an data examples leads to more precise and detailed descriptions.

Figure~\ref{fig:length} compares the length of the original descriptions with those generated by multiple ablation of our system. For version 1 (V1), we also tested various model configurations, including increasing the number of parameters from 8 billions to 70 billions and adjusting the model temperature. 
However, we observed no significant differences between these configurations. 
Therefore, we proceeded with the Llama 3 8B at 0.5 temperature for the rest of the experiments. 
Regardless of the model hyperparameters, we consistently generate descriptions that were longer than the original user-written descriptions, with the latter being around 800 characters (250 tokens) and ours being more than double in length, around 2000 characters (500 tokens). 
The difference in length also highlights the need to encode the generated descriptions in a way that accounts for their longer context length. Most off-of-the-shelf sequence encoding models have a limit of 250 or 512 tokens, which can result in truncated embeddings for longer descriptions. 
To address this, we chunk the generated descriptions into smaller segments and compute a cumulative embedding as described in Section~\ref{sec:RAG-indexing}.

\begin{figure*}[!ht]
\hfill
\subfloat[]
{\includegraphics[width=.6\linewidth]{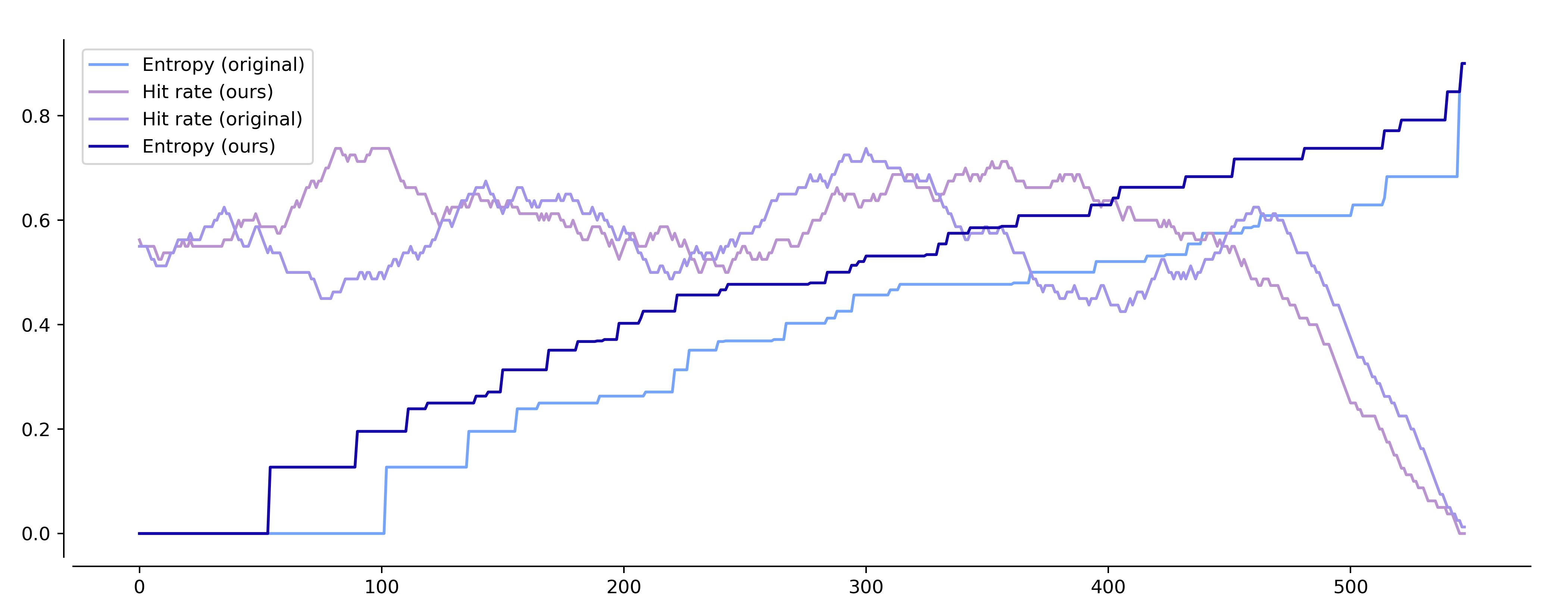}}
\hfill
\subfloat[]
{\includegraphics[width=0.3\linewidth]{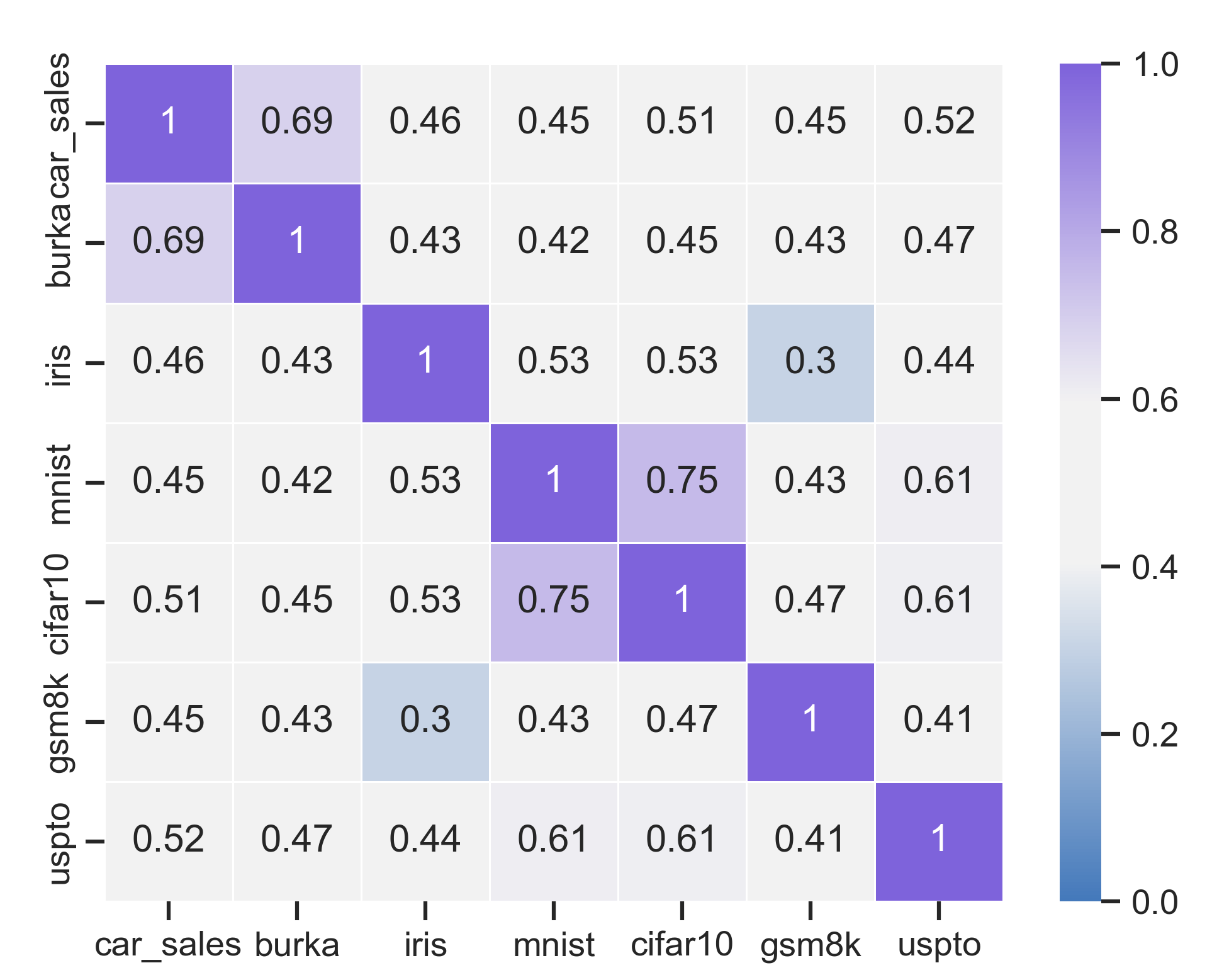}}
\hfill
\caption{(a)Hit rate against normalized retrieval entropy of RAG based on the descriptions generated by our approach (V3-Llama-8B-0.5T) and on the original descriptions. For both measures, the higher the better. (b) Cosine similarity between pairs of vectors used by our system for the retrieval of popular HF dataset benchmarks. A value of $1$ corresponds to high-similarity in the vector space, indicating that the two datasets are related, while a value of $0$ indicates poor similarity.}
\label{fig:retrieval-eval}
\end{figure*}

\subsection{Redundancy of original descriptions}
Figure~\ref{fig:length}c highlights the redundancy present in the original Zenodo descriptions when compared to the corresponding papers. 
The plot displays the distribution of pairwise similarities between each Zenodo record and its respective publication. 
As observed, the distribution is heavily skewed toward high similarity values, indicating that the descriptions on Zenodo can be largely reconstructed from the paper alone.
To further validate this observation, we examined the record descriptions for all entries with similarity scores ranging from $0.8$ to $1.0$. 
We found that these high-similarity descriptions were either direct copies of the paper's abstract (resulting in a similarity score of $1$) or replicas of the paper's title.

In contrast, the generated descriptions exhibit a broader distribution of similarity scores relative to the papers. 
This is encouraging, as our goal is to create descriptions that complement the information presented in the papers by primarily leveraging the underlying data content.
\subsection{Retrieval assessment}
Following the evaluation in Section~\ref{sec:evaluation}, we generate $15$ questions for each record with an available associated paper, starting from the paper content, reaching a total number of $548$ questions. Examples of the questions are reported in the Appendix \ref{appendix:q_and_a_retrieval}.

We build two vector stores to compare the efficacy of the generated descriptions. The first vector store was built by using our system descriptions generated with the V3 Llama 3 8B model at temperature 0.5. The second vector store, used for baseline comparison, was built using the original descriptions in Zenodo.
Apart from the top-1 accuracy which was comparable to the original descriptions, at $38$\% against $39$\%, our method outperformed the original descriptions in all the other retrieval evaluation tasks. The increase on the top-5 accuracy is of $2$\% points ($53$\% of our system against $51$\% of the baseline) and of $4$\% points on the top-10 accuracy ($60$\% against $56$\%).

An analysis of the hit rate against the retrieval entropy is illustrated in Figure~\ref{fig:retrieval-eval} (a), where the questions are sorted for increasing entropy. 
Our method shows higher entropy already within the first hundred questions. 
This means that the retrieval results were more diverse than those obtained with the baseline vector store. 
Within the first $150$ questions sorted by entropy, our method reported $12$\% higher top-10 accuracy than the baseline (with $63$\% against $51$\%) respectively. 
This shows that the retrieval based on our description is not only more accurate on relatively easy questions, but that it is also more diversified, allowing to find additional related datasets. 

\subsection{Extension to other datasets}
\label{sec:res-hf}
We extend the analysis to HF datasets, following the criteria described in~\ref{evaluation-hf}. We evaluate the vector similarity between the representations of the Zenodo records and classic machine learning datasets such as
MNIST~\cite{lecun1998mnist}, CIFAR10~\cite{cifar10}, the Iris dataset~\cite{iris_53}, the Car prices datasets~\cite{cars_dataset}, the Burka-1999 Czech Financial Dataset (CFD)~\cite{burka1000czech} and the USPTO large-scale benchmark dataset of annotated molecule images~\cite{Lowe2017chemical}. 
These datasets were chosen among popular benchmarks from the HF datasets catalog, and were included in the automated analysis to generate the data curation reports. 
Figure~\ref{fig:retrieval-eval} (b)     illustrates the cosine similarity between the vectors obtained for each dataset. 
As expected, the similarity values are relatively low for datasets that represent diverse type of data and tasks, whereas they are high for the CIFAR10 and MNIST pair at $0.75$ and for the CFD and car sales dataset at $0.69$. 
Moreover, the similarity between vectors representing pair of files in the \textit{nccr-catalysis} community is high, averaging at $0.88$ for file-level reports belonging to the same record.

\subsection{Re-usability of the data analysis results}
\label{section:data_synth}
\begin{figure}[!h]
\subfloat[]
{
\includegraphics[width=0.8\linewidth]{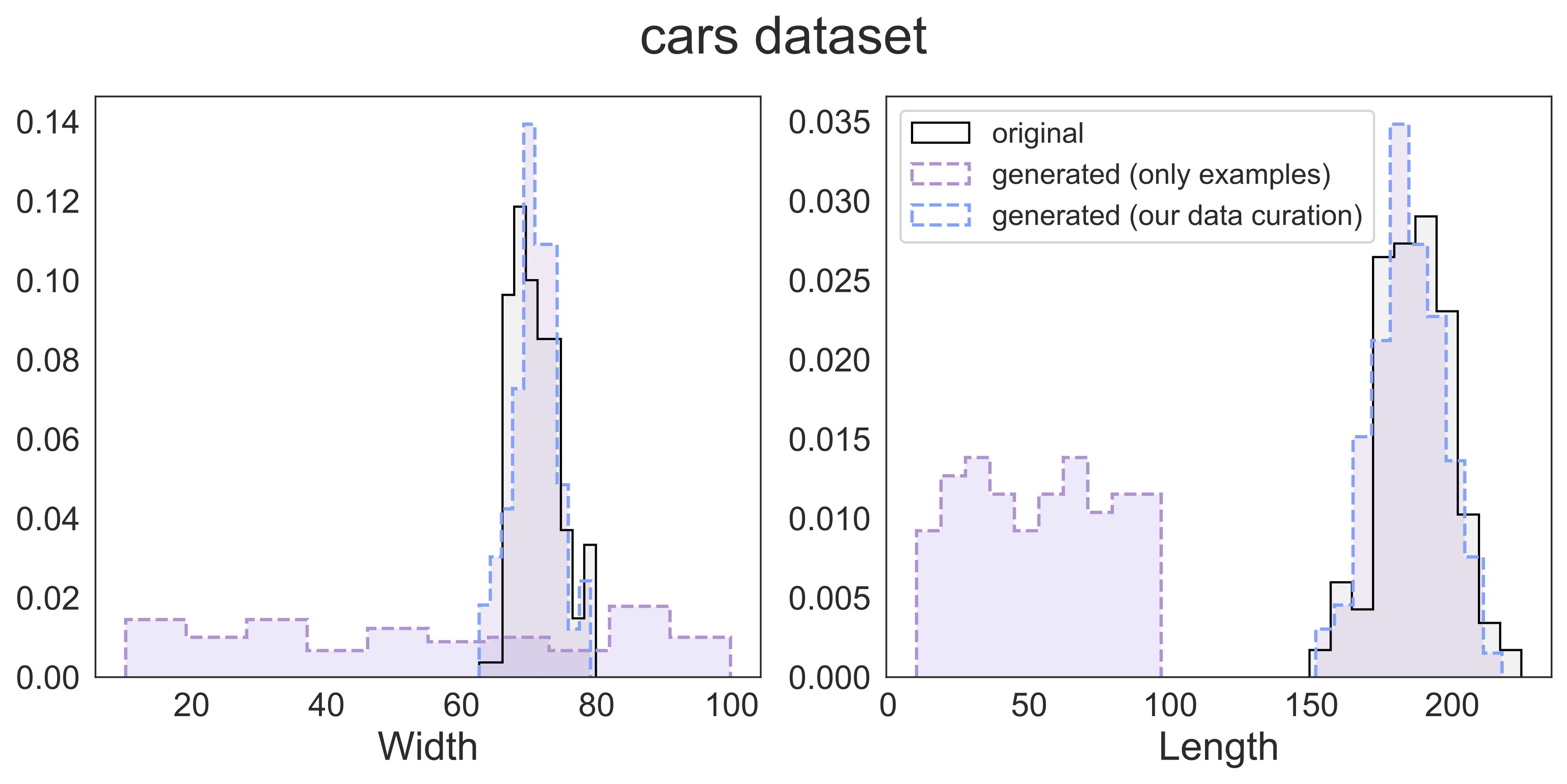}
}
\vfill
\subfloat[]
{
\includegraphics[width=0.8\linewidth]{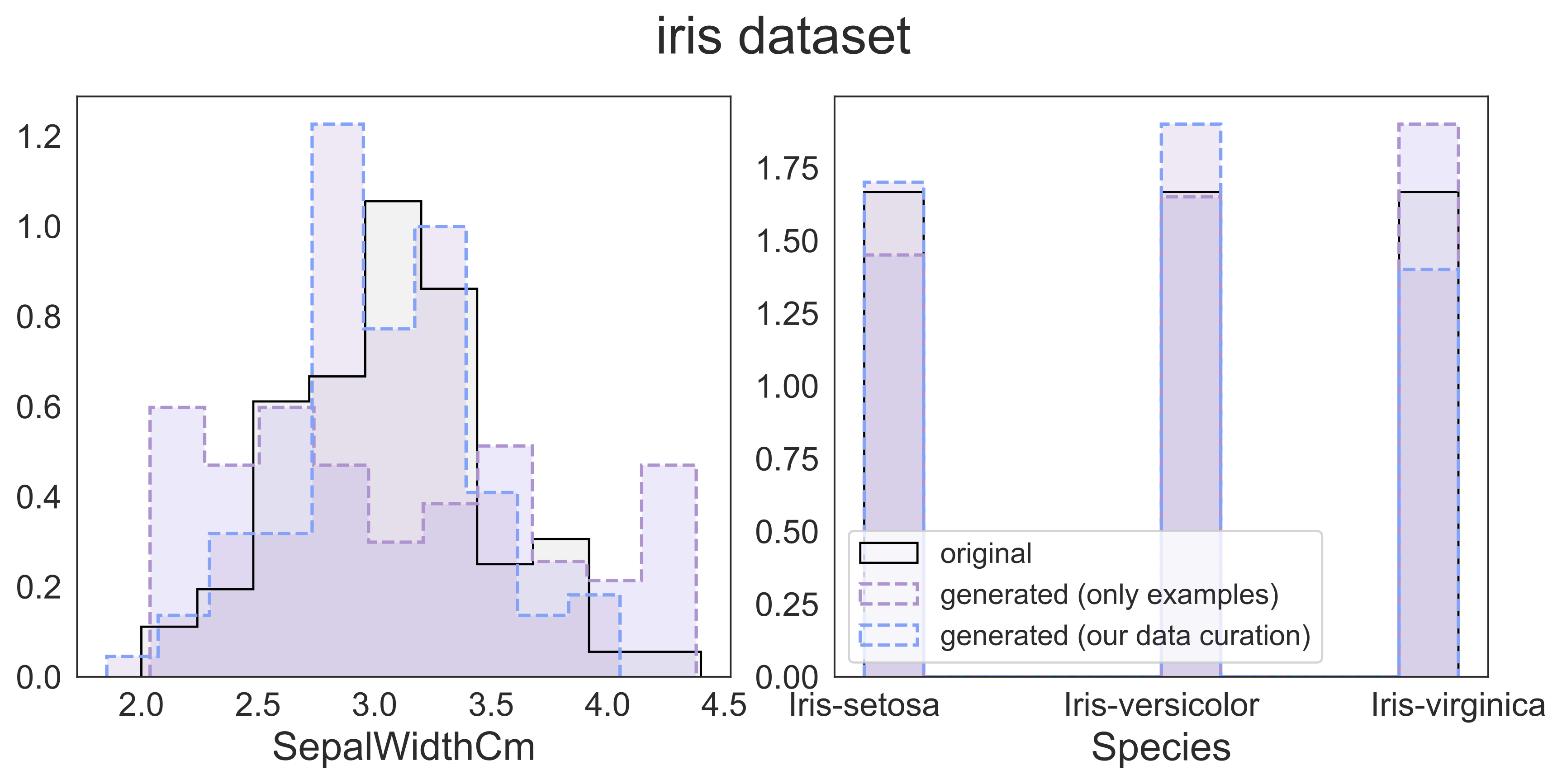}
}
\caption{Qualitative comparison of synthetic data generated with or without our metadata curation. a) Car Sales b) Iris Flowers }
\label{fig:datasynth}
\end{figure}

In this section, we present qualitative and quantitative results for yet another task where using our metadata curation offers an advantage compared to existing work. 
For these experiments we focus on two popular tabular datasets available on HF datasets, namely Car Sales~\cite{cars_dataset} and Iris Flowers~\cite{iris_53}.
A comparison between the distributions of the real and synthetic data is illustrated in Figure~\ref{fig:datasynth}.
For a quantitative assessment, in Table \ref{tab:data_synth_metrics} we calculate the percentage of overlap for each feature between the original histogram and each of the generated ones, having fixed range and number of bins.
These results show that our data curation leads to substantially more realistic synthetic data samples. 

\begin{table}[!h]
\begin{tabular}{p{2.7cm}p{2.4cm}p{2.1cm}}
\hline
Dataset {[}feature{]}   & Area overlap\newline(our data curation) & Area overlap\newline(only examples)\\ \hline
Cars {[}Width{]}        & \textbf{95}\%                                       & 16\%                                   \\ 
Cars {[}Length{]}       & \textbf{94}\%                                       & 0\%                                    \\ 
Iris {[}SepalWidthCm{]} & \textbf{88}\%                                       & 62\%                                   \\ 
\hline
\end{tabular}
\caption{Percentage of area overlap between the original dataset distribution and the generated ones for different features of the Iris and Cars datasets. Comparisons are column-wise and the higher the better ($\uparrow$).}
\label{tab:data_synth_metrics}
\end{table}

\section{Related Work}
In this section, we contextualize the novelty of our approach as opposed to existing work, drawing from the main concepts and approaches in the related literature. We cover three main aspects that are related to this work, namely the concept of data curation through gathering metadata, the use of intelligent agents to understand unstructured data, and, finally, the related work on RAG for dataset discovery. 

Funding institutions and researchers collaborated to establish the FAIR (Findability, Accessibility, Interoperability and Reusability) principles for data sharing in the context of open science, aiming to accelerate dataset discovery and enhance the reproducibility of research outcomes~\cite{hodson2018turning}. 
The principles emphasize the importance of packaging data with related information like source code, configuration instructions and metadata about the experiments~\cite{hodson2018turning,marsolek2023datacuration,nust2021datadiscovery}, highlighting that effective data curation can greatly enhance dataset discovery and reuse, even when datasets are not publicly shared or accessible~\cite{sheridan2021datadiscovery}.
However, the concept of metadata is versatile, being declined in schemas that are tailored to each application. 
Some works derive metadata from the project title and keywords~\cite{necasky2022datadiscovery,sakaji2021datadiscovery}, while others use the associated publication~\cite{basmah2021datadiscovery} or a description of the dataset's features~\cite{koesten2018datadiscovery}.
Combining the features with textual summaries, in particular, was shown to prioritize the retrieval of relevant results by excluding irrelevant ones~\cite{koesten2018datadiscovery}.  
Numerous studies support the idea that including contextual information facilitates the "sense making" process when exploring a  dataset~\cite{hemphill2022properties,sheridan2021datadiscovery,gebru2021datasheets,koch2021reduced,marsolek2023datacuration,marchionini2005datadiscovery,koesten2018datadiscovery}.
In addition to that,~\cite{bernhauer2022datadiscovery} proposed that chaining the search from one dataset to related ones can complement the dataset search. 
In this work, we build on top of these studies by developing a more comprehensive metadata schema that incorporates the results of the automated data analysis performed by intelligent agents. 

Intelligent agents have gained traction in recent research for analyzing unstructured data~\cite{cheng2023gpt,peng2023study,tayebi2024large,ansari2024agent}. 
Particularly, GPT-like models were shown to outperform junior and intern data analysts in solving analysis tasks and providing correct answers.
Researchers further analyzed the utility of GPT in addressing clinical questions, comparing the performance to that of expert physicians~\cite{peng2023study,tayebi2024large}. 
However, these works motivate an unsafe use of data sharing, as sensitive data may be exposed by individual file uploads to third-party models. 
In contrast, our approach involves processing entire sections of online data repositories locally, thereby minimizing data sharing and enhancing privacy.
Moreover, the organization of the results in a vector space facilitates the retrieval of related datasets, complementing single file analysis as proposed in~\cite{bernhauer2022datadiscovery}. 

A related concept is RAG, introduced by Meta to combine vector databases with language generation~\cite{lewis2020retrieval}. 
This methodology has proven effective in scientific document understanding, with RAG outperforming students in scientific question answering tasks~\cite{auer2024docling,lala2023paperqa, skarlinski2024language}. In the context of dataset discovery, RAG was applied by~\cite{hayashi2024datadiscovery} to improve metadata-based data exploration, focusing on fields such as dataset summary, variable names and keywords. In contrast to our approach, however, the metadata in their work was already available, with no automated extraction involved. 

\section{Conclusion}
The retrieval and organization of various datasets related to a specific research topic is a significant challenge faced by machine learning researchers. 
As the volume of unstructured digital data grows, sorting, cataloging, and indexing this information becomes increasingly critical for ensuring traceability and effective data management.
In this paper, we proposed a solution that demonstrates how intelligent agents, combined with data analysis tools and online dataset repositories, can substantially reduce the costs of data curation. 
Specifically, we show that this approach can minimize curation costs to levels comparable to the inference costs of small transformer models. 
Through experimentation with a niche use case in scientific discovery, we highlight how our method enhances the curation and organization of experimental data, improving data retrieval outcomes and providing valuable analytical insights. 
These improvements can promote dataset reusability and foster new discoveries. 

Moreover, when applied to guide synthetic data generation the data analysis insights generated by our system proved effective, showing that it could even aid researchers in tasks such as data augmentation. 
Ultimately, we have shown that the integration with large data repositories for machine learning research such as HF datasets is possible and that it can provide insteresting gateways for transforming experimental data into machine-learning-ready datasets. 

Currently, a key limitation of our system lies in its storage complexity. 
The full datasets must be downloaded on disk to perform the data analysis. 
Scaling up this method to petabytes of data would require advanced distributed storage solutions, such as cloud-based architectures (e.g., AWS S3 or Google Cloud Storage) combined with efficient data streaming and processing frameworks like Apache Spark or Dask. 
A simpler and more practical alternative, however, is to perform the analysis in batches, reducing the need for extensive storage while maintaining scalability.
An even more efficient solution would involve integrating this systems directly in the public data repositories during the the data upload process, where researchers could validate and refine the generated content before the dataset publication. 

Whether applied prior to data publication or afterward, solutions like the one proposed here are crucial for facilitating  dataset discovery and re-use, thereby maximising the value of the existing publicly available datasets. 
Solutions such as the one proposed in our work address the need for robust frameworks to facilitate data usability, which is becoming ever more critical as governmental funding agencies increasingly mandate the open publication of datasets generated through publicly funded research.
Shared datasets often lack sufficient curation, as human effort alone, particularly from researchers without expertise in data management, proves insufficient to meet the standards of true reusability.

Our system addresses this challenge by enabling efficient dataset processing and integration with lightweight transfomer models capable of running locally. This empowers researchers to analyze and explore datasets within their areas of interest, reducing the reliance on ‘easy-choice’ datasets and mitigating biases observed in prior studies~\cite{koch2021reduced}.
Moving forward, the development and adoption of similar frameworks will be essential to transform publicly available datasets into genuinely reusable resources, unlocking their full potential for accelerating scientific discovery across disciplines.
\begin{acks}
This work was supported by the Swiss National Science Foundation (SNSF) through the National Centre of Competence in Research for Catalysis (NCCR-Catalysis, grant number 180544).
We thank Marvin Alberts and Oliver Schilter for the initial insights on data retrieval during the design stage of the experiments. Moreover, we thank Antonio Focubierta Rodríguez for the constructive feedback provided to improve the clarity and readability of our manuscript. 
Finally, we thank Matteo Manica for providing suggestions that turned relevant for the shaping of the methods, starting from the integration with LangChain and the use of code generation to generate the synthetic examples. 
\end{acks}

\bibliographystyle{ACM-Reference-Format}
\bibliography{main}

\pagebreak
\appendix



\section{Qualitative assessment}
\label{appendix:qualitative-assessment}
Here we provide more examples of the data analysis report generated by our system, focusing on the HF datasets mentioned in Section~\ref{sec:res-hf}. 
For simplicity, we only report the textual information, although the system also generates a pickle file including data statistics such as individual feature distributions, feature correlations and dataset statistics. 
\\
\\
\fbox{\begin{minipage}{28em}
\textbf{Dataset}: USPTO \\
\textbf{Description}:\\
This dataset contains metadata from the United States Patent and Trademark Office (USPTO) for approximately 30,000 patent applications. Each record represents a single patent application and includes three main features: 'filename', 'image', and 'mol'.
[...]\\
The dataset contains image features.
Based on the captions, it appears that the input types are descriptions of molecular structures, specifically:
\begin{itemize}
    \item The structural structure of the aminos (likely referring to amino acids, the building blocks of proteins)
    \item The structural structure of the benze molecule (referring to the benzene molecule, a common organic compound)
 \item The structural structure of a molecule (a general term that could refer to any type of molecule). 
\end{itemize}
These captions suggest that the input types are likely to be chemical structures or molecular formulas, which would be used to analyze and understand the properties and behavior of these molecules.\\
\textbf{Domain}: Chemistry \\
\textbf{Keywords}: Chemical Structure Images, Patent Documents, Molecular Data, Image Processing, Patent Analysis \\
\end{minipage}}
\\
\\
\fbox{\begin{minipage}{28em}
\textbf{Dataset}: MNIST \\
\textbf{Description}:\\
The MNIST dataset contains a collection of images and corresponding labels. The images are described in the "image" feature, which is a non-numeric categorical variable. The labels are represented by the "label" feature, which is a numeric categorical variable with 10 possible values (0-9).\\
Based on the captions, it appears that the input types are a mix of categorical and numerical data. The captions describe various logos, which suggests that the input types may include image or string data. The presence of numerical data is hinted at by the caption "the number 7 in the sky", which implies that the dataset may include integer or floating-point values. Overall, the input types may include a combination of categorical (logos, text) and numerical (numbers) data.
[...]\\
\textbf{Domain}: Image\\
\textbf{Keywords}: Image Classification Dataset, Categorical and Numerical Data, Logo Recognition, Object Detection, Multimodal Data 
\end{minipage}}

\section{Retrieval assessment}
\label{appendix:q_and_a_retrieval}
We report examples of the generated questions for the retrieval task. 
The questions were generated by the Llama-8B language model with 0.5 temperature. 
Question examples:
\newcounter{question}
\setcounter{question}{0}
\Que{Can you find a dataset to investigate the effect of Fe impurities on the performance of a hybrid CoPc@MWCNT catalyst during CO2 electrolysis, including information on current transients and gas chromatography results?}

\Que{Which dataset contains data related to the structural alterations of silver nanoparticles (Ag NPs) during CO2 electrolysis in a zero-gap gas-flow configuration?}

\Que{What are the inventory datasets used for life cycle assessment of Fischer-Tropsch Diesel production from CO2 and H2 sources, including direct air capture and point source coal power plant, as well as hydrogen production from biomass and polymer electrolyte water electrolysis?}

\Que{Can you find a dataset on the application of operando laser scattering to study the behavior of complex electrode architectures in electrochemical systems?}

\Que{What are the specific reaction conditions (e.g. temperature, pressure, solvent) used for the Suzuki-Miyaura cross-coupling reaction?}

\Que{How do the thermal activation and surface properties of Sr0.8Ca0.2FeO3-d relate to the overall performance of the oxygen carrier material, and what are the exact values of the performance metrics (e.g. oxygen storage capacity, relaxation times) at each cycle?}

\Que{What are the exact X-ray powder diffraction patterns of Sr0.8Ca0.2FeO3-d before and after thermal treatment in a CO2-free atmosphere?}

\section{Re-usability of data analysis results}
In this section we provide supplemental information to the \ref{section:data_synth} Results subsection, namely the code generation prompts in~\ref{appendix:prompt_data_synth}, the generation parameters in~\ref{appendix:params_data_synth} and the generation scripts in~\ref{appendix:scripts_data_synth}.
\subsection{Code generation prompts}
\label{appendix:prompt_data_synth}
Prompt used for generation of code which at its turn will generate synthetic data using our metadata curation method.
\begin{tcolorbox}
"""\\
You are an agent designed to write and execute python code to generate synthetic data from a query.
You have access to a python REPL, which you can use to execute python code.
If you get an error, debug your code and try again until there are no errors. \\
Generate 100 synthetic data samples about the \{'subject'\} dataset.
Examples here \{'examples'\}. \\
Now, generate python code that creates a pandas dataframe where each column is sampled \textit{according to their statistical information contained in \{'metadata\_stats'\}}.\\
Save the pandas dataframe in a csv file \{'output\_file'\}.\\
"""
\end{tcolorbox}
The sentence in italics is the only part of the prompt that contains explicit instructions to use the metadata report, in the case of generating synthetic data based only on examples this is the only sentence that varies to ``as in the examples above.''

\subsection{Generation parameters}
\label{appendix:params_data_synth}
\begin{tcolorbox}
\begin{verbatim}
model_name: meta-llama/llama-3-1-70b-instruct
decoding_method: "greedy" 
min_new_tokens: 20
temperature: int = 1
stop_sequences: ["stop"]
random_seed:= 42
repetition_penalty:  1.0
top_p: 1
truncate_input_tokens:  0
\end{verbatim}
\end{tcolorbox}

\subsection{Data generation scripts}
\label{appendix:scripts_data_synth}
\lstset{
    language=python,
    tabsize=3,
    caption=Python code generated by the LLM using our metadata curation for the Cars dataset,
    label=code:sample,
    frame=shadowbox,
    rulesepcolor=\color{gray},
    xleftmargin=20pt,
    framexleftmargin=15pt,
    keywordstyle=\color{blue}\bf,
    commentstyle=\color{gray},
    stringstyle=\color{red},
    numbers=left,
    numberstyle=\tiny,
    numbersep=5pt,
    breaklines=true,
    showstringspaces=false,
    basicstyle=\footnotesize,
    emph={str},emphstyle={\color{magenta}}}
\begin{figure*}[t]
\begin{lstlisting}[language=Python, breaklines=true]
import pandas as pd
import numpy as np

np.random.seed(0)

# Define the stats for each column
stats = {
    'year_resale_value': {'mean': 18.07297520661157, 'std': 11.453384099225005, 'min': 5.16, 'max': 67.55},
    'Price_in_thousands': {'mean': 27.390754838709682, 'std': 14.351653189238547, 'min': 9.235, 'max': 85.5},
    'Width': {'mean': 71.15, 'std': 3.451871862459368, 'min': 62.6, 'max': 79.9},
    'Length': {'mean': 187.34358974358975, 'std': 13.431754284516822, 'min': 149.4, 'max': 224.5},
    'Vehicle_type': {'top': 'Passenger', 'freq': 116, 'unique': 2}
}

# Generate the data
data = {
    'year_resale_value': np.clip(np.random.normal(stats['year_resale_value']['mean'], stats['year_resale_value']['std'], 100), stats['year_resale_value']['min'], stats['year_resale_value']['max']),
    'Price_in_thousands': np.clip(np.random.normal(stats['Price_in_thousands']['mean'], stats['Price_in_thousands']['std'], 100), stats['Price_in_thousands']['min'], stats['Price_in_thousands']['max']),
    'Width': np.clip(np.random.normal(stats['Width']['mean'], stats['Width']['std'], 100), stats['Width']['min'], stats['Width']['max']),
    'Length': np.clip(np.random.normal(stats['Length']['mean'], stats['Length']['std'], 100), stats['Length']['min'], stats['Length']['max']),
    'Vehicle_type': np.random.choice(['Passenger', 'Car'], 100, p=[stats['Vehicle_type']['freq'] / 157, 1 - stats['Vehicle_type']['freq'] / 157])
}

# Create the dataframe
df = pd.DataFrame(data)

df.to_csv('/output_folder/generated_code_stats.csv', index=False)
\end{lstlisting}
\end{figure*}

\lstset{
    language=python,
    tabsize=3,
    caption=Python code generated by the LLM using examples only the Cars dataset,
    label=code:sample2s,
    frame=shadowbox,
    rulesepcolor=\color{gray},
    xleftmargin=20pt,
    framexleftmargin=15pt,
    keywordstyle=\color{blue}\bf,
    commentstyle=\color{gray},
    stringstyle=\color{red},
    numbers=left,
    numberstyle=\tiny,
    numbersep=5pt,
    breaklines=true,
    showstringspaces=false,
    basicstyle=\footnotesize,
    emph={str},emphstyle={\color{magenta}}}
\begin{figure*}[b]
\begin{lstlisting}[language=Python, breaklines=true]

import pandas as pd
import numpy as np

# Define the columns
columns = ['Manufacturer', 'Model', 'Sales_in_thousands', 'year_resale_value', 'Vehicle_type', 'Price_in_thousands', 'Engine_size', 'Horsepower', 'Wheelbase', 'Width', 'Length', 'Curb_weight', 'Fuel_capacity', 'Fuel_efficiency', 'Latest_Launch', 'Power_perf_factor']

# Define the data types
data_types = [str, str, float, float, str, float, float, float, float, float, float, float, float, float, str, float]

# Generate synthetic data
data = []
for _ in range(100):
    row = []
    for i, data_type in enumerate(data_types):
        if data_type == str:
            if i == 0:  # Manufacturer
                row.append(np.random.choice(['Nissan', 'Pontiac', 'BMW', 'Volvo', 'Plymouth']))
            elif i == 1:  # Model
                row.append(np.random.choice(['Quest', 'Montana', '528i', 'S40', 'Voyager']))
            elif i == 4:  # Vehicle_type
                row.append(np.random.choice(['Car', 'Passenger']))
            else:  # Latest_Launch
                row.append(f'{np.random.randint(1, 13)}/{np.random.randint(1, 28)}/{np.random.randint(2010, 2013)}')
        else:
            if i == 3:  # year_resale_value
                row.append(np.random.uniform(10, 40) if np.random.rand() < 0.5 else np.nan)
            else:
                row.append(np.random.uniform(10, 100))
    data.append(row)

# Create a pandas DataFrame
df = pd.DataFrame(data, columns=columns)


df.to_csv('/output_folder/generated_code_no_stats.csv', index=False)
\end{lstlisting}
\end{figure*}

\end{document}